\documentclass[preprint,12pt]{elsarticle}

\usepackage{amssymb}

\usepackage{url}

\journal{Physics Letters B}

\begin{document}

\begin{frontmatter}



\title{A New Measurement of Kaonic Hydrogen X rays}


\author[lnf]{M.~Bazzi}
\author[victoria]{G.~Beer}
\author[milano]{L.~Bombelli}
\author[lnf,ifin]{A.M.~Bragadireanu}
\author[smi]{M.~Cargnelli\corref{cor}}
\ead{michael.cargnelli@oeaw.ac.at}
\author[lnf]{G.~Corradi}
\author[lnf]{C.~Curceanu (Petrascu)}
\author[lnf]{A.~d'Uffizi}
\author[milano]{C.~Fiorini}
\author[milano]{T.~Frizzi}
\author[roma]{F.~Ghio}
\author[roma]{B.~Girolami}
\author[lnf]{C.~Guaraldo}
\author[ut]{R.S.~Hayano}
\author[lnf,ifin]{M.~Iliescu}
\author[smi]{T.~Ishiwatari}
\author[riken]{M.~Iwasaki}
\author[smi,tum]{P.~Kienle}
\author[lnf]{P.~Levi~Sandri}
\author[milano]{A.~Longoni}
\author[lnf]{V.~Lucherini}
\author[smi]{J.~Marton}
\author[lnf]{S.~Okada\corref{cor}}
\ead{shinji.okada@lnf.infn.it}
\author[lnf]{D.~Pietreanu}
\author[ifin]{T.~Ponta}
\author[lnf]{A.~Rizzo}
\author[lnf]{A.~Romero~Vidal}
\author[lnf]{A.~Scordo}
\author[ut]{H.~Shi}
\author[lnf,ifin]{D.L.~Sirghi}
\author[lnf,ifin]{F.~Sirghi}
\author[ut]{H.~Tatsuno}
\author[ifin]{A.~Tudorache}
\author[ifin]{V.~Tudorache}
\author[lnf]{O.~Vazquez Doce}
\author[smi]{E.~Widmann}
\author[smi]{J.~Zmeskal}
\cortext[cor]{Corresponding authors.}
 
\address[lnf]{INFN, Laboratori Nazionali di Frascati, Frascati (Roma), Italy}
\address[victoria]{Dep. of Phys. and Astro., Univ. of Victoria, Victoria B.C., Canada}
\address[milano]{Politecnico di Milano, Sez. di Elettronica, Milano, Italy}
\address[ifin]{IFIN-HH, Magurele, Bucharest, Romania}
\address[smi]{Stefan-Meyer-Institut f$\ddot{\mbox{u}}$r
 subatomare Physik, Vienna, Austria}
\address[roma]{INFN Sez. di Roma I and Inst. Superiore di Sanita, Roma, Italy}
\address[ut]{Univ. of Tokyo, Tokyo, Japan}
\address[riken]{RIKEN, The Inst. of Phys. and Chem. Research, Saitama, Japan}
\address[tum]{Tech. Univ. M$\ddot{\mbox{u}}$nchen, Physik Dep., Garching, Germany}

\begin{abstract}
 The $\overline{K}N$ system at threshold is a sensitive testing ground for 
 low energy QCD,
 especially for the explicit chiral symmetry breaking.
 Therefore, we have measured the $K$-series x rays of kaonic hydrogen atoms
 at the DA$\Phi$NE electron-positron collider
 of Laboratori Nazionali di Frascati,
 and have determined the most precise values of
 the strong-interaction energy-level
 shift and width of the $1s$ atomic state.
 As x-ray detectors,
 we used large-area silicon drift detectors
 having excellent energy and timing resolution,
 which were developed especially for the
 SIDDHARTA experiment.
 The shift and width were determined to be
 $\epsilon_{1s} = -283 \pm 36 \mbox{(stat)} \pm 6 \mbox{(syst)}$ eV
 and $\Gamma_{1s} = 541 \pm 89 \mbox{(stat)} \pm 22 \mbox{(syst)}$ eV,
 respectively.
 The new values will provide vital constraints
 on the theoretical description of the low-energy $\overline{K}N$ interaction.
\end{abstract}


\end{frontmatter}


\section{Introduction}
\label{sec:intro}

Low energy phenomena in strong interactions are described by
effective field theories which
contain appropriate degrees of freedom to describe physical phenomena
occurring at the nucleon-meson scale.
Chiral perturbation theory
was extremely successful in describing systems like pionic atoms,
however it is not directly applicable for the kaonic systems. This is
due to the presence of resonances like the $\Lambda$(1405), only
slightly below the $K^-p$ reaction threshold of 1432 MeV. Instead,
non-perturbative coupled-channel techniques are used.
These calculations generate the $\Lambda$(1405)
dynamically as a $\overline{K}N$ quasibound state and as a resonance in
the $\pi$$\Sigma$ channel. A general feature of the theory in this field
is that it relies heavily on input from experimental data.

The measurement of the strong-interaction induced energy-level shift and width
of the kaonic-hydrogen $1s$ atomic state provides direct
information on the $\overline{K}N$ $s$-wave interaction at $K^-p$ threshold.
Kaonic-hydrogen x-ray data are therefore important
for theories of the $\overline{K}N$ system,
together with the experimental data of low-energy $K^-p$ scattering,
$\pi \Sigma$ mass spectra, and $K^-p$ threshold decay ratios.
These studies allow the investigation of
chiral SU(3) dynamics in low-energy QCD and the role of explicit chiral
symmetry-breaking due to the relatively large strange quark mass.
These data are also strongly related to recent hot topics
-- the structure of the $\Lambda(1405)$ resonance 
($e.g.,$ \cite{twopole,hyodo,hj})
and the deeply bound kaonic systems ($e.g.,$ \cite{AY,FINUDA,KEK,DISTO}).
Recent progress in this field is summarized in \cite{ECT}.

The shift and width are deduced
from the spectroscopy of the $K$-series kaonic-hydrogen x rays.
The first distinct peaks of the kaonic-hydrogen x rays
were observed by the KEK-PS E228 group \cite{kpx}
following the absorption of a stopped $K^-$
within a gaseous hydrogen target
using Si(Li) detectors.
The observed repulsive shift
was consistent with the analysis of
the low energy $\overline{K}N$ scattering data,
resolving the long-standing sign discrepancy
generated by old experiments
\cite{J_D_Davies_1979,M_Izycki_1980,P_M_Bird_1983}.

The most recent values were reported by the DEAR experiment \cite{dear},
in which the errors were reduced by a factor of 2
when compared with those of E228 \cite{kpx}.

Using the results obtained by DEAR, theoretical studies have been
performed with possible higher order contributions using several
models
\cite{correct,PRL2005,nissler,oller,oller2,bor,oller3,revai,cieply,shevchenko,oset}.
However, the question still remains
that most of them had difficulties in explaining
all the experimental results in a consistent way.
See also \cite{NPA2010,PPNP}.
  
Here we report on results
based on the x-ray detection technique recently developed by the
SIDDHARTA (Silicon Drift Detector for Hadronic Atom Research by Timing
Application) collaboration,
using the microsecond timing and excellent energy resolution of
large area silicon drift detectors (SDDs) \cite{NIM-tomo}.
This technique
reduced the large x-ray background coming from beam losses
and improved the signal-to-background ratio
by more than a factor of 10 with respect to 
the corresponding DEAR ratio of about 1/100.

\section{Experiment}
\label{sec:exp}

The SIDDHARTA experiment was performed at
the DA$\rm{\Phi}$NE electron-positron collider
at the Laboratori Nazionali di Frascati of INFN.
The $\phi$-resonances produced decay
into back-to-back $K^+K^-$ pairs emitted with a branching ratio
of about 49 \%.
The monochromatic low-energy kaons ($\sim$ 16 MeV of kinetic energy)
are stopped efficiently in a gaseous target to
produce kaonic hydrogen atoms.
It is essential to use a gaseous target for
the measurements since the x-ray yields quickly decrease towards higher
density due to the Stark mixing effect.
Therefore, a cryogenic gaseous hydrogen target was used
at typical values of a pressure 0.1 MPa and a temperature 23 K,
resulting in a density of $1.3 \times 10^{-3}$ g/cm$^3$
of the isotopically pure hydrogen.

\begin{figure}[t]
 \begin{center}
  \includegraphics*[width=7.5cm]{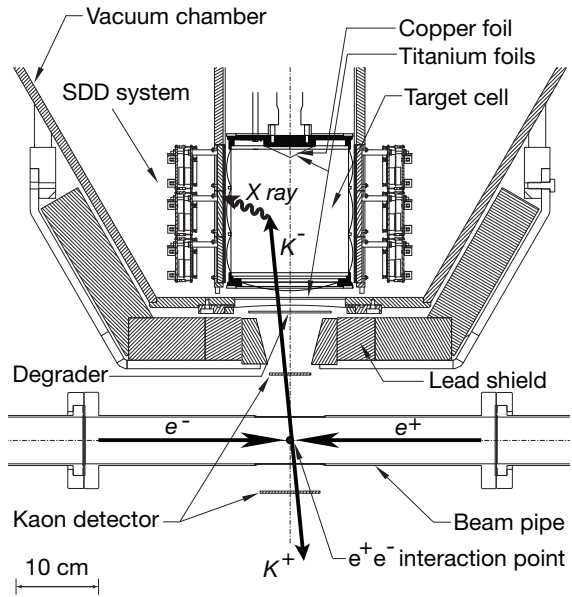}
  \caption{A schematic side view of the SIDDHARTA setup installed at
  the e$^+$e$^-$ interaction point of DA$\rm{\Phi}$NE.}
  \label{fig:Setup}
 \end{center}
\end{figure}

Figure \ref{fig:Setup} shows the SIDDHARTA setup.
To detect the back-to-back correlated $K^+K^-$ pairs
from $\phi$ decay,
two plastic scintillation counters (1.5 mm thick), called the kaon detector,
were mounted above and below the $e^+e^-$ interaction point
where the $\phi$ resonance is produced.
A kaon trigger was defined by the coincidence of
the two scintillators.
Minimum ionizing particles (MIPs) coming from beam losses
were highly suppressed
by setting a high pulse-height threshold in the kaon detector
-- the slow kaons deposit much more energy in the scintillators
than the faster MIPs.

To obtain a uniform distribution of $K^-$ momenta entering the gaseous
target, a shaped degrader made of mylar foils with a thickness ranging
from 100 to 800 $\mu$m was placed as shown in Fig. \ref{fig:Setup},
to correct for a slight momentum boost of the $\phi$
resulting from the 55 mrad $e^+ e^-$ crossing angle.

The cylindrical target cell,
13.7 cm in diameter and 15.5 cm high,
was located just above the degrader
inside the vacuum chamber.
The lateral wall and the bottom window
were made of Kapton Polyimide film
of 75 $\mu$m and 50 $\mu$m thickness.

The SDDs,
used to detect the kaonic-atom x rays, were developed within
a European research project devoted to this experiment \cite{NIM-tomo}.
Each of the 144 SDDs used in the apparatus has an area of 1 cm$^2$ and a
thickness of 450 $\mu$m.
The SDDs, operated at a temperature of $\sim$170 K, had an energy
resolution of
183 eV (FWHM) at 8 keV
(a factor of 2 better than Si(Li) detectors
used in E228 \cite{kpx}) and timing resolution below 1 $\mu$sec
in contrast to the CCD detectors used in DEAR \cite{DEAR-CCD}
which had no timing capability.
Using the coincidence between $K^+K^-$ pairs and x rays measured by SDDs,
the main source of background coming from beam losses was highly suppressed.

To test our experimental technique and optimize the degrader thickness,
we repeatedly changed the target
filling to helium gas and measured
the $L$-transitions of kaonic $^4$He.
Due to the high yield of this kaonic atom x-ray transition,
one day of measurement was sufficient for each check.

The physics results of the strong-interaction $2p$-level shifts
of kaonic $^3$He and kaonic $^4$He atoms are available in our recent
publications \cite{SIDT2011,SIDT2009}.

In addition, we have performed the first-ever exploratory measurement
of kaonic-deuterium $K$-series x rays with the same experimental setup.
In the kaonic-hydrogen analysis, 
it turned out to be essential to use the
kaonic-deuterium spectrum to quantify the kaonic background x-ray lines
-- originating from kaons captured in heavier elements such as carbon,
nitrogen and oxygen contained in organic construction materials --
which overlap the kaonic-hydrogen signal.

Data were accumulated over six months in 2009 with integrated
luminosities of $\sim$ 340 pb$^{-1}$ for the hydrogen
and $\sim$ 100 pb$^{-1}$ for the deuterium measurement.

\section{Data analysis}
\label{sec:ana}

\begin{figure}[t]
 \begin{center}
  \includegraphics*[width=7.5cm]{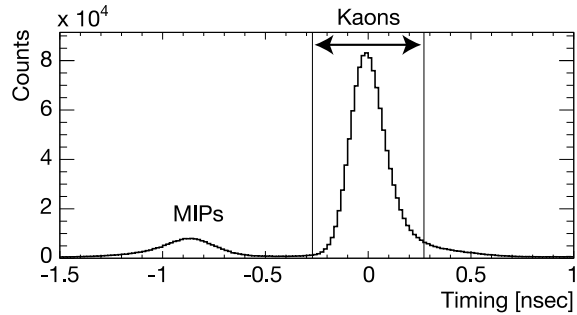}
  \caption{Kaon identification
  using timing of the coincidence signals in the
  kaon detector with respect to the RF signal of $\sim$ 368.7 MHz
  from DA$\Phi$NE.
  }
  \label{fig:KID}
 \end{center}
\end{figure}

The data acquisition system records
the signal amplitudes seen by the 144 detectors
along with the global time information.
Whenever a kaon trigger occurred,
the time difference between the x ray and kaon was recorded as well as
the time correlations between the signals on each of the scintillators
and the DA$\Phi$NE bunch frequency.
From these data, the time-of-flight information of the kaon detector,
the position of the hit detector and rates of the SDDs, rate of kaon
production, etc., could be extracted in the off-line analysis.

The timing distribution
of the coincidence signals in the kaon detector 
with respect to the RF signal from DA$\Phi$NE
shows clearly that
kaon events can be separated from MIPs
by setting a time gate as indicated by arrows in Fig. \ref{fig:KID}.

\begin{figure}[t]
 \begin{center}
  \includegraphics*[width=7.5cm]{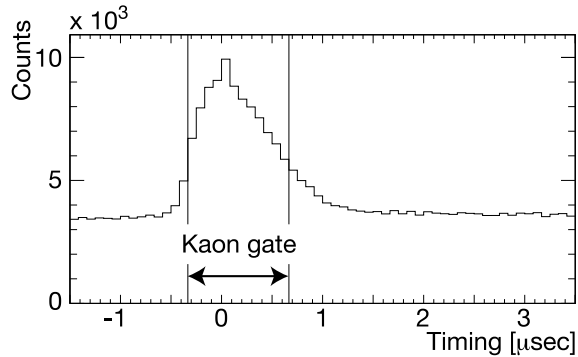}
  \caption{Time difference spectrum between kaon arrival
  and x-ray detection for $K^-$ triggered events of hydrogen data,
  where a time-walk correction was applied.
  }
  \label{fig:TimeEne}
 \end{center}
\end{figure}

The time difference between kaon arrival and x-ray detection
for hydrogen data is shown in Fig. \ref{fig:TimeEne}.
The peak represents correlation between x rays and kaons,
while the flat underlying structure is from
uncorrelated accidental background.
A typical width of the time-correlation, after a time-walk correction,
was about 800 ns (FWHM) which reflected the drift-time distribution
of the electrons in the SDD.

In order to sum up the individual SDDs, 
the energy calibration of each single SDD
was performed by periodic measurements
of fluorescence x-ray lines from titanium and copper foils, excited by an
x-ray tube, with the $e^+e^-$ beams in kaon production mode.
A remote-controlled system moved the kaon detector out and the x-ray
tube in for these calibration measurements, once every $\sim$ 4 hours.

The refined {\it in-situ} calibration 
in gain (energy) and resolution (response
shape) of the summed spectrum of all SDDs was
obtained using titanium, copper, and gold fluorescence lines 
excited by the uncorrelated background without trigger
(see \cite{SIDT2011,SIDT2009} for more details),
and also using the kaonic carbon lines from wall stops
in the triggered mode.

\begin{figure}[t]
 \begin{center}
  \includegraphics*[width=7.5cm]{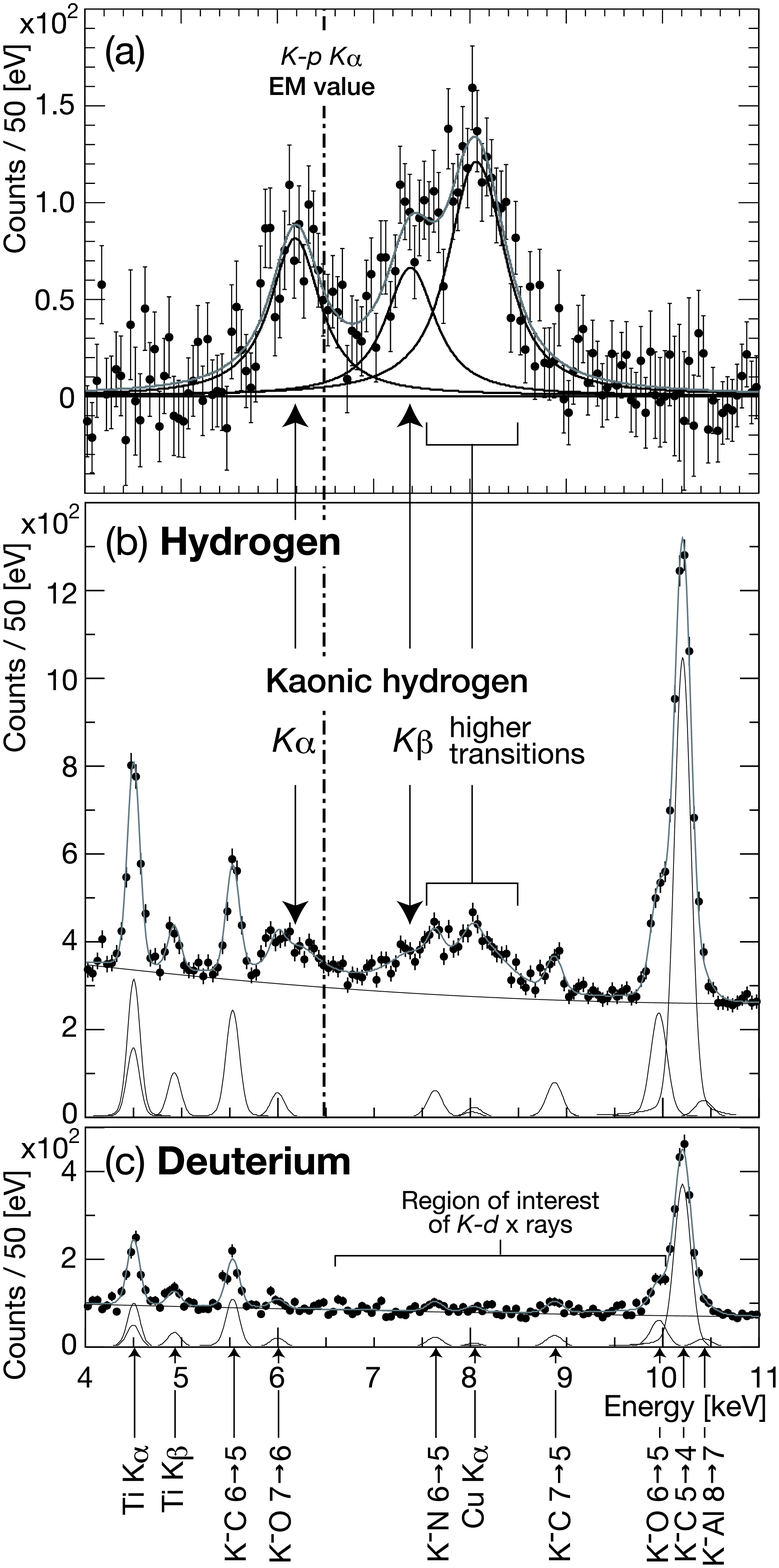}
  \caption{A global simultaneous fit result of
  the x-ray energy spectra of hydrogen and deuterium data.
  (a) Residuals of the measured kaonic-hydrogen x-ray spectrum
  after subtraction of the fitted background,
  clearly displaying the kaonic-hydrogen $K$-series transitions.
  The fit components of the $K^-p$ transitions are also shown,
  where the sum of the function is drawn
  for the higher transitions (greater than $K\beta$).
  (b)(c) Measured energy spectra with the fit lines.
  Fit components of the background x-ray lines
  and a continuous background are also shown.
  The dot-dashed vertical line indicates
  the EM value of the kaonic-hydrogen $K\alpha$ energy.
  (Note that the fluorescence $K\alpha$ line consists of
  $K\alpha1$ and $K\alpha2$ lines, both of which are shown.)
  }
 \label{fig:SpecFit}
 \end{center}
\end{figure}

Figure \ref{fig:SpecFit} shows
the final kaonic hydrogen and deuterium x-ray energy spectra.
$K$-series x rays of kaonic hydrogen were clearly observed
while those for kaonic deuterium were not visible.
This appears to be consistent with the theoretical expectation of lower
x-ray yield and greater transition width for deuterium
($e.g.,$ \cite{T_Koike_1996}).

The vertical dot-dashed line in Fig.\ \ref{fig:SpecFit}
indicates the x-ray energy of kaonic-hydrogen $K\alpha$
calculated using only the electro-magnetic interaction (EM).
Comparing the kaonic-hydrogen $K\alpha$ peak
and the EM value,
a repulsive shift (negative $\epsilon_{1s}$)
of the kaonic-hydrogen $1s$-energy level is easily seen.

Many other lines from kaonic-atom x rays and characteristic x rays
were detected in both spectra as indicated
with arrows in the figure.
These kaonic-atom lines result from
high-$n$ x-ray transitions
of kaons stopped in the target-cell wall made of Kapton
(C$_{22}$H$_{10}$O$_5$N$_2$)
and its support frames made of aluminium.
There are also characteristic x rays from titanium and copper
foils installed for x-ray energy calibration.

We performed a global simultaneous fit
of the hydrogen and deuterium spectra.
The intensities of the three background x-ray lines
overlapping with the kaonic-hydrogen signals
(kaonic oxygen 7-6, kaonic nitrogen 6-5, and copper $K\alpha$)
were determined using both spectra and a normalization factor defined by
the ratio of the high-statistics kaonic-carbon 5-4 peak in the $K^-p$
and $K^-d$ spectra.
Figure \ref{fig:SpecFit}, (b) and (c) show the fit result
with the components of the background x-ray lines
and a continuous background;
(a) shows the residuals of the measured kaonic-hydrogen x-ray spectrum
after subtraction of the fitted background,
clearly displaying the kaonic-hydrogen $K$-series transitions.

The response function
of the SDD detectors
was found
to contain a slight deviation from a pure Gaussian shape, which could
influence the determination of the strong-interaction width of the
kaonic-hydrogen x-ray lines.
The deviation from the pure Gaussian response was treated in two different 
ways to test systematic effects.

The first analysis used satellite peaks on the left and
right flanks. The left flank is generally found in
silicon detectors and is defined as a function having a feature decreasing
exponentially in intensity towards lower energy \cite{ResponseFunc}.
The right flank could be interpreted
as an effect of pile-up events and was defined
as a Gaussian \cite{e570}.

In the second analysis, the correction of the ``ideal'' response was done
by convoluting the Gauss function with a Lorentzian (producing symmetric
tails) and additionally with an exponential low energy tail.

In both analyses, the kaonic-hydrogen lines were represented by Lorentz functions
convoluted with the detector response function, where the Lorentz width
corresponds to the strong-interaction broadening of the $1s$ state.
The continuous background was represented by a quadratic polynomial function.

The region of interest of $K^-d$ x rays is illustrated in
Fig. \ref{fig:SpecFit} (c).
With the realistic assumption
of one order of magnitude lower intensities
than that of kaonic hydrogen \cite{T_Koike_1996}
and predicted values
of shift ($-0.3$ - $-1.0$ keV) and width ($\sim$ 1 keV)
\cite{barrett99, wycech04, meissner06},
the influence of a possible kaonic-deuterium component
on the kaonic-hydrogen shift and width values
was found to be negligible.

In the kaonic-hydrogen spectrum, the higher transitions to the $1s$
level, $K\gamma$ and above, produce an important
contribution to the total intensity.
The relative intensities of these lines, however, are only poorly known
from cascade calculations and the free fit cannot accurately distinguish
between them, since their relative energy differences are smaller than
their width, as seen in Fig.\ \ref{fig:SpecFit}.
As a result, fitting all the transitions at once,
leads to large errors on the shift and width of the $1s$ level.
To minimize the influence of the higher transitions, we adopted the
following iterative fitting procedure.
In a first step, we performed a free fit of all the transitions,
with the energy differences between the kaonic-hydrogen
lines fixed by their EM differences, since the shifts and the widths of
the levels higher than $1s$ are negligible.
In a second step, we fixed the energies and widths of the higher
transitions to the values found in the first step, and fitted leaving
free all intensities and the common shift and width for $K\alpha$ and
$K\beta$, which are well resolved transitions.
With the new values for shift and width we repeated the described
procedure until the values for the shift and width converged, meaning
that all $K$-lines had the same values for their shift and width,
as it should be.

We performed two independent analyses,
where the event selection, the calibration method, the fit range
and the detector-response function (as described above)
were chosen differently.
The comparison of the shift and width
values gives a direct measurement of the systematic error
from the use of differing procedures.
The resulting shift values were consistent with each other
within 1 eV,
however the width differed by $\sim$ 40 eV
which comes mainly from the use of different detector-response functions.
For shift and width we quote here the mean value of the two analyses
and take into account the difference as
one of the sources of the systematic error.

As a result, the $1s$-level shift $\epsilon_{1s}$ and width
$\Gamma_{1s}$ of kaonic hydrogen were determined to be
\begin{eqnarray*}
\epsilon_{1s} = -283 \pm 36 \rm{(stat)} \pm 6 \rm{(syst) ~eV} \\
\rm{and}~~ \Gamma_{1s} = 541 \pm 89 \rm{(stat)} \pm 22 \rm{(syst) ~eV},
\end{eqnarray*}
respectively,
where the first error is statistical and the second is systematic.
The quoted systematic error is a quadratic summation
of the contributions from the ambiguities due to
the SDD gain shift,
the SDD response function,
the ADC linearity,
the low-energy tail of the kaonic-hydrogen higher transitions,
the energy resolution,
and the procedural dependence shown by independent analysis.

If we could fix the intensity pattern of all transitions,
which would be possible if more accurate cascade calculations existed,
the statistical error would be better than $\pm$ 25 eV for shift
and $\pm$ 55 eV for width.

\section{Conclusion}
\label{sec:concl}

In conclusion, we have performed the most precise measurement of
the $K$-series x rays of kaonic hydrogen atoms.
This was made possible by the use of new triggerable x-ray detectors,
SDDs, developed in the framework of the SIDDHARTA project,
which lead to a much improved energy and time resolution
over the past experiments \cite{kpx,dear}
and much lower background in comparison with the DEAR experiment.

\begin{figure}[t]
 \begin{center}
  \includegraphics*[width=7.5cm]{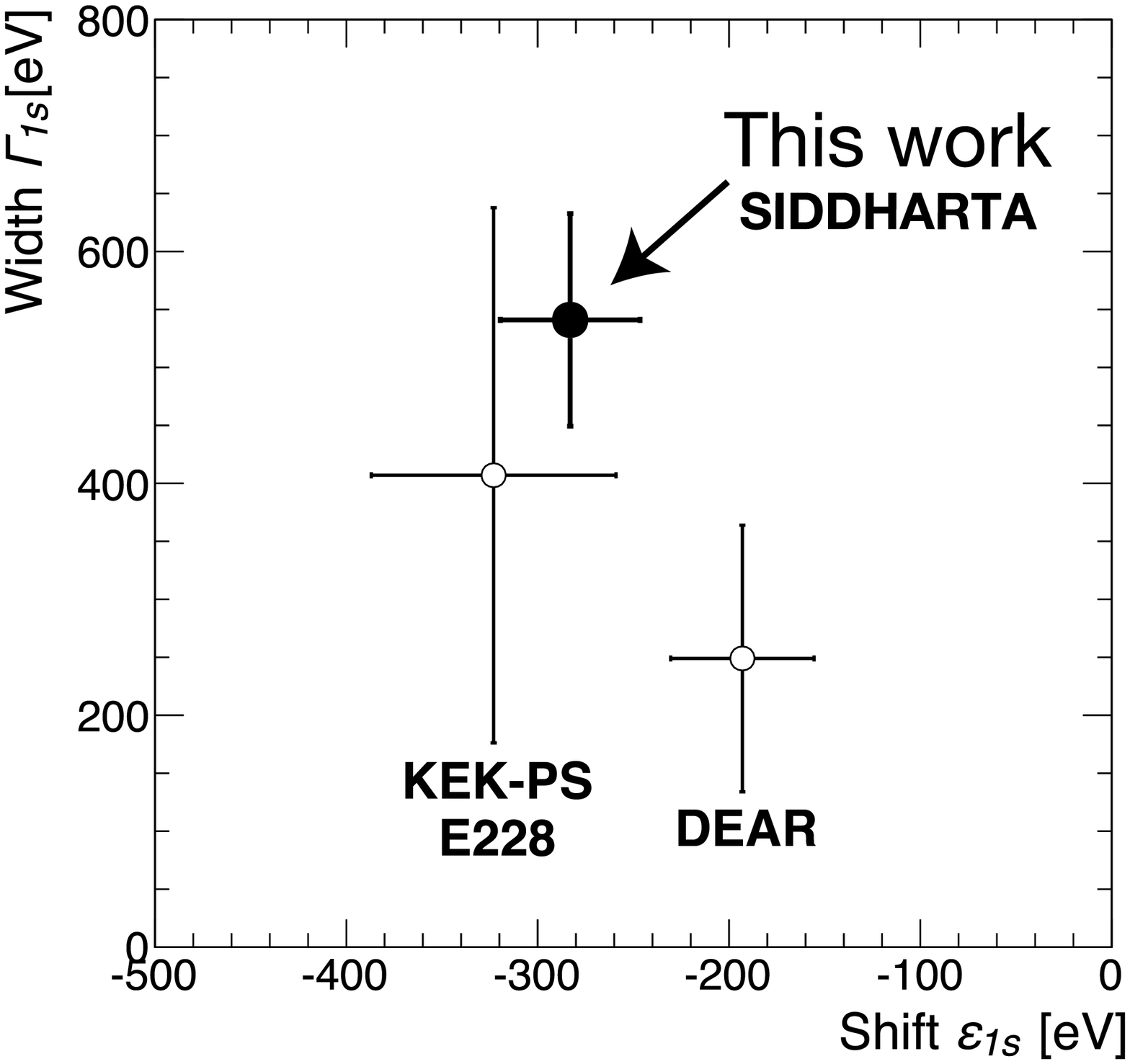}
  \caption{Comparison of experimental results for the
  strong-interaction $1s$-energy-level shift
  and width of kaonic hydrogen,
  KEK-PS E228 \cite{kpx} and DEAR \cite{dear}.
  The error bars correspond to quadratically added statistical and systematic
  errors.
  }
  \label{fig:Comparison}
 \end{center}
\end{figure}

The strong-interaction $1s$-energy level shift and width of
kaonic hydrogen are
plotted in Fig. \ref{fig:Comparison}
along with the results of the previous two measurements,
E228 \cite{kpx} and DEAR \cite{dear}.

Our determination of the shift and width
does provide new constraints
on theories,
having reached a quality which will demand refined calculations
of the low-energy $\overline{K}N$ interaction.

For further study of the $\overline{K}N$ interaction,
it is essential to measure
the kaonic-deuterium $K$-series x rays
to disentangle the isoscalar and isovector scattering lengths.
The present result combined with deuterium data to be collected
in the SIDDHARTA-2 experiment \cite{sidd2} will provide invaluable
knowledge about the behavior of low-energy QCD in the strangeness sector.

\section*{Acknowledgments}
\label{sec:ack}

We thank C. Capoccia, B. Dulach, and D. Tagnani from LNF- INFN; and
H. Schneider, L. Stohwasser, and D. St$\ddot{\mbox{u}}$ckler
from Stefan-Meyer-Institut,
for their fundamental contribution in designing and building the
SIDDHARTA setup.
We thank as well the DA$\Phi$NE staff for the excellent working
conditions and permanent support.
Part of this work was supported by
HadronPhysics I3 FP6 European Community program,
Contract No. RII3-CT-2004-506078;
the European Community-Research Infrastructure Integrating
Activity ``Study of Strongly Interacting Matter''
(HadronPhysics2, Grant Agreement No. 227431)
under the Seventh Framework Programme of EU;
Austrian Federal Ministry of Science
and Research BMBWK 650962/0001 VI/2/2009;
Romanian National Authority for Scientific Research,
Contract No. 2-CeX 06-11-11/2006;
and the Grant-in-Aid for Specially Promoted Research (20002003), MEXT, Japan.








\end{document}